\documentclass[twocolumn,twoside,aps,superscriptaddress,prl,showpacs]{revtex4}
\usepackage{graphicx}
\usepackage{dcolumn}
\usepackage{amssymb,amsmath}
\usepackage{citesort}
\usepackage{amssymb}
\usepackage{ifpdf}

\DeclareMathOperator{\erfc}{erfc}

\def\be{\begin{eqnarray}}
\def\ee{\end{eqnarray}}
\begin{document}
\bibliographystyle{apsrev}

\title{Giant bubble pinch--off}
\author{Raymond Bergmann$^1$, Devaraj van der Meer$^1$, Mark Stijnman$^1$,
Marijn Sandtke$^1$, Andrea Prosperetti$^{1,2}$, and Detlef Lohse}
\affiliation{ Physics of Fluids Group and J.M. Burgers Centre for
Fluid Dynamics, University of Twente, P.O. Box 217,
7500 AE Enschede, The Netherlands\\
$\mathit{^2}$ Department of Mechanical Engineering, The
Johns-Hopkins University, Baltimore, Maryland 21218, USA}

\begin{abstract}
Self-similarity has been the paradigmatic picture for the pinch-off
of a drop. Here we will show through high-speed imaging and boundary
integral simulations that the inverse problem, the pinch-off of an
air bubble in water, is {\it not} self-similar in a strict sense: A
disk is quickly pulled through a water surface, leading to a giant,
cylindrical void which after collapse creates an upward and a
downward jet. Only in the limiting case of large Froude number the
neck radius $h$ scales  as $h(-\log h)^{1/4} \propto \tau^{1/2}$,
the purely inertial scaling. For any finite Froude number the
collapse is slower, and a second length-scale, the curvature of the
void, comes into play. Both length-scales are found to exhibit
power-law scaling in time, but with different exponents depending on
the Froude number, signaling the non-universality of the bubble
pinch-off.
\end{abstract}


\maketitle

The pinch--off of a liquid drop is a prime example of a hydrodynamic
singularity and has been studied extensively in recent years
\cite{ber93,bre94,sto94,egg97,day98,zha99b,coh99}. It has become
paradigmatic for self--similar behavior in fluid dynamics: After
appropriate rescaling, the shapes of the pinching neck at different
times can be superimposed onto a single shape
\cite{egg97,day98,zha99b,coh99}. With the exception of some
pioneering work \cite{lon91,ogu93}, the {\it inverse} problem of the
collapse of a gas--filled neck surrounded by a liquid has not
attracted much attention until very recently, with the analysis of
the pinch--off of a bubble rising from a needle and the break--up of
a gas bubble in a straining flow
\cite{dos03,lep03,sur04,bur05,gor05}. The time--evolution of these
collapsing gas--filled necks is found to follow a power law. If the
dynamics near the singularity are solely governed by liquid inertia,
then the radius of the neck $h$ expressed in the time $\tau$
remaining until collapse scales as $h\propto\tau^{1/2}$
\cite{lon91,ogu93,dos03,bur05}, or, with a logarithmic correction,
as $h(-\log{h})^{1/4}\propto \tau^{1/2}$ \cite{gor05}. Deviations
from this exponent of $1/2$ are reported to occur only due to the
inclusion of other effects. The collapse may be slowed down by
viscosity ($h\propto\tau$ \cite{dos03,sur04,bur05}) or surface
tension ($h\propto\tau^{2/3}$ \cite{lep03}), or accelerated by the
inertia of the gas flowing inside the neck, leading to
$h\propto\tau^{1/3}$ \cite{gor05}.

In this paper we focus on another example of this ``inverse
pinch-off'', namely the violent collapse of the void created at a
fluid surface by the impact of an object. Here we find exponents
which deviate substantially from 1/2, even though the dynamics are
shown to be purely governed by liquid inertia, without significant
contributions from the effects mentioned above. The self-similar
behavior $h(-\log{h})^{1/4}\propto \tau^{1/2}$ appears to hold only
in the asymptotic regime of very high impact velocities.

\begin{figure*}
\includegraphics[scale=3.5]{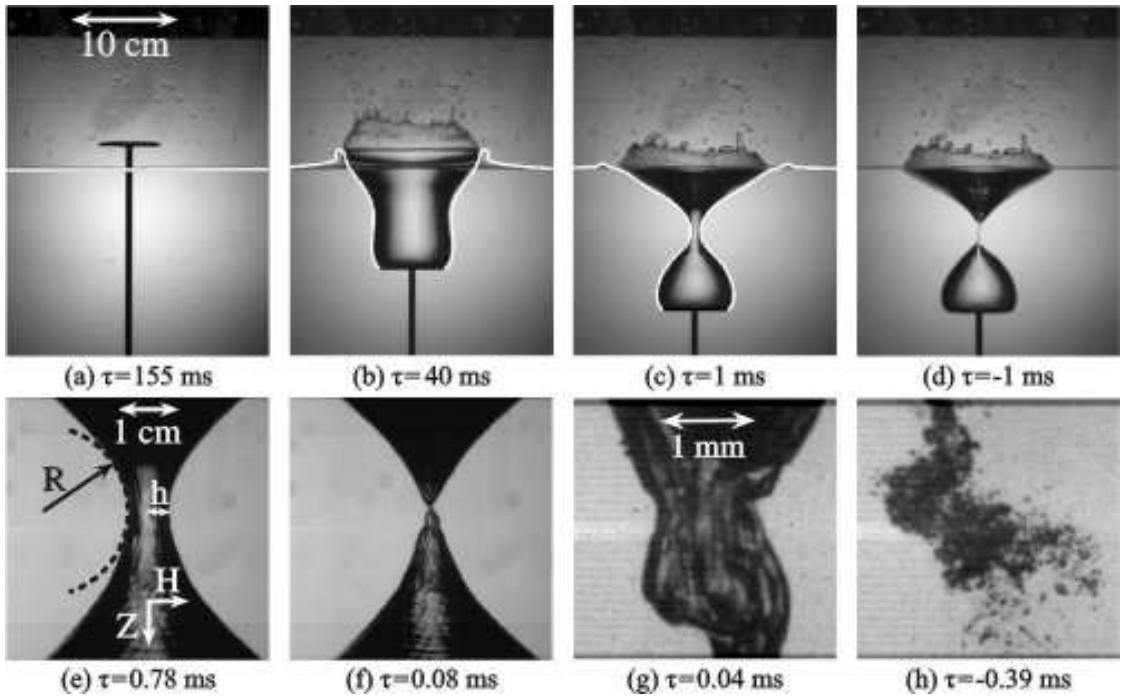}
\caption{\small Snapshots of the formation and collapse of a surface
void for the plunger experiment: A linear motor pulls down a disk of
radius $h_{disk}$ = 30 mm through the water surface at a constant
velocity $V = 1.0$ m/s (Fr = 3.4). (a-d) The collapse of the void
imaged at a 1000 frames per second. The white lines (overlay) are
the void profiles obtained from boundary integral simulations with
the same initial condition, and without the use of any free
parameter. (e-f) Details of the collapse imaged at 12800 frames per
second. (g-h) Details of the collapse imaged at 48000 frames per
second. Note that the field of view decreases with increasing frame
rate. In the very last stages of the collapse (f-g) there is a
Kevin-Helmholtz instability that complicates the determination of
the time of collapse. Immediately after the collapse air is
entrapped, both in the form of a large bubble above disk (d) and as
a cloud of microbubbles at the collapse point (h).}
\label{snapshots}
\end{figure*}

In our experiment, a linear motor is used to drag metal disks with
radii $h_{disk}$ between 10 and 40 mm through an air/water interface
with a well-controlled, constant velocity $V$ between 0.5 and 3 m/s
(see Fig.~\ref{snapshots}a). A rod running through a seal in the
bottom of a large water tank connects the disk with the linear
motor. This arrangement generates giant voids in a very controlled
fashion. The advantage of this setup is that the velocity is a {\it
control parameter} and not the response of the object to the fluid
forces upon impact. Secondly, due to the large scale of the
experiment, viscosity and surface tension play a negligible role
\footnote{Viscosity and surface tension effects are quantified by
the magnitude of the Reynolds ($\mathrm{Re}$) and Weber
($\mathrm{We}$) numbers, which are considerable ($> 10^2$) during
the pinch-off process. This holds when they are defined globally,
i.e., with respect to the impact velocity and the disk radius
(Re~$=h_{disk}V/\nu$ and We~$=h_{disk}V^2 \rho/ \sigma $), but also
when they are defined locally using the neck radius and velocity at
a specific time (Re~$=h \dot{h}/\nu$ and We~$=h\dot{h}^2 \rho/
\sigma$).}. Therefore the only important dimensionless parameter is
the Froude number $\mathrm{Fr}\,=\,V^2/(h_{disk}g)$, the ratio of
kinetic to gravitational energy, which ranges from $0.6$ to $90$.
The large scale of the experiment is also advantageous for the
observation of details during the impact and collapse process, which
is imaged with digital high-speed cameras with frame rates up to
$100,000$ frames per second.

A typical series of events is seen in Fig.~\ref{snapshots}a-d. The
impact of the disk creates an axisymmetric void which first expands
until the hydrostatic pressure drives the walls inward. The inward
moving walls collide and cause a pinch-off at some depth below the
undisturbed free surface. The energy focusing of this violent
collapse creates a strong pressure spike on the axis of symmetry
which releases itself in a downward and an upward jet
\cite{lat00,det04}. The latter reaches heights exceeding $1.5$ m for
the higher impact speeds in this experiment. It is this dominating
role of inertia that makes our system different from other pinch-off
processes in the literature. At higher recording speeds the
pinch-off can be investigated in more detail as in
Fig.~\ref{snapshots}e-h. There is a clear loss of both azimuthal and
axial symmetry in Figs.~\ref{snapshots}f and \ref{snapshots}g, which
can be attributed to a combination of the same convergence effect
that causes an instability in a collapsing bubble
\cite{ple77,hao99,hil96}, and a Kelvin-Helmholtz instability due to
the rapid air flow in the neck. The latter increases with increasing
Froude number and limits the range of our experiments. Another
factor which limits the Froude number range is the so-called surface
seal, in which the void closes at the water surface as the
crown-like splash is entrained by the air flowing into the expanding
void \cite{bir57,gau98}. This process, which occurs at large Froude
numbers, changes the pinch-off considerably since in this case the
gas pressure inside the void differs appreciably from that of the
ambient air.

In view of these experimental limitations, we performed numerical
simulations using a boundary integral method based on potential
theory {\it without} ambient gas. There is an excellent agreement
between the numerical calculations and the experiments, as seen in
Fig.~\ref{snapshots}a-c. Here, the numerical void profiles (the
solid white lines) coincide very well with the experimental profiles
in the pinch-off region without the use of any adjustable parameter,
either in space or in time.

\begin{figure*}[top!]
\includegraphics[scale=2]{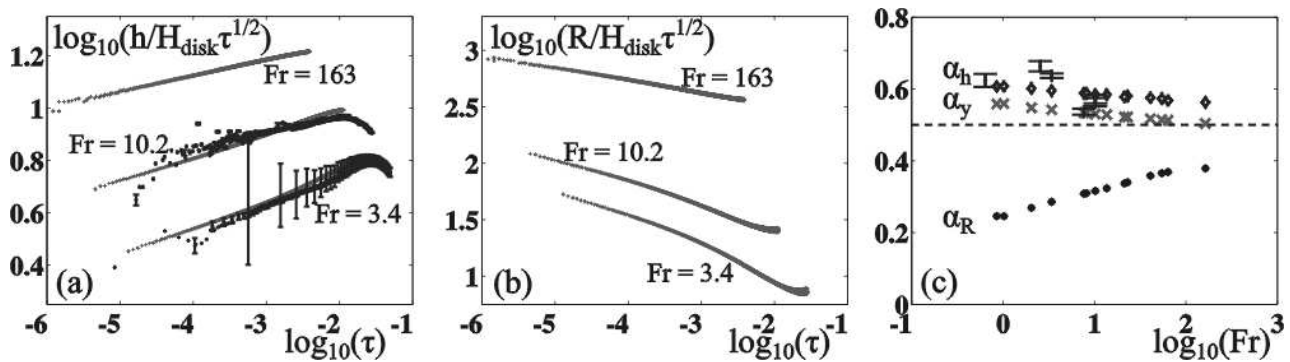}
\caption{\small (a) The radius of the void at the depth of closure
$h$, compensated with $\tau^{1/2}$, as a function of the time $\tau$
remaining until collapse in a doubly logarithmic plot, for three
different values of the Froude number $\mathrm{Fr}$. Experiment
(blue symbols) and numerical simulations (red symbols) are seen to
agree very well for $\mathrm{Fr}$ = 3.4 and 10.1. For $\mathrm{Fr}$
= 163 only numerical data are presented, because for this Froude
number experiments are hindered by the surface seal (see text). The
error bars, indicating the error in the experimental data, are
usually small, but occasionally become very large for frames very
close to the collapse time. (b) Doubly logarithmic plot of the
radius of curvature of the void profile $R$ compensated with
$\tau^{1/2}$ as a function of $\tau$ for the numerical simulations
of (a). Both $h$ and $R$ are well described by power laws for up to
four orders of magnitude in $\tau$. (c) Power-law exponents
$\alpha_h$ for the radius of the void at closure depth $h(\tau)$,
$\alpha_y$ for the radius of the void including the logarithmic
correction $y(\tau) = h(-\log{h})^{1/4}$, and $\alpha_R$ for the
radius of curvature of the void at closure depth $R$, all as a
function of the Froude number.} \label{plots}
\end{figure*}

To further quantify the pinch-off process, we now turn to the time
evolution of the neck radius $h(\tau)$, measured at the depth at
which the void eventually closes. Because both length and time
scales become very small close to collapse, it is not feasible to
experimentally observe the collapse with only one high-speed camera
recording \footnote{We image the pinch-off process over four orders
of magnitude in time and two in space. As the field of view of the
camera corresponds to $10^3$ pixels, this would leave only 10 pixels
for the last stage of the collapse. Moreover, the hole sequence
should then be imaged at a frame rate corresponding to the smallest
timescale ($10 \mu$s), i.e., 100 kHz requiring at least 10 GB of
fast storage capacity, greatly exceeding the physical capabilities
of our cameras.}. Due to the reproducibility of the experiment, we
overcame this difficulty by matching several data sets imaged at
different frame rates, increasingly magnifying the region around the
pinch-off. Figure~\ref{plots}a contains a doubly logarithmic plot of
$h(\tau)$ (compensated with $\tau^{1/2}$) for both the high-speed
imaged experiments and the numerical calculations, again showing
excellent agreement for different Froude numbers. In this graph, a
straight line corresponds to the power law behavior $h =h_0
\tau^{\alpha_h}$. The exponent $\alpha_h$ is plotted as a function
of $\mathrm{Fr}$ in Fig.~\ref{plots}c. Clearly, there are large
deviations from the suggested behavior $\alpha_h=1/2$. Can these be
explained by a logarithmic correction as proposed in \cite{gor05}?

Let us first establish the origin of this logarithmic correction in
our system. Near the neck, the flow induced by the collapsing void
looks very much like that of a collapsing cylinder, while it must
look like that of a sink, plus its image in the free surface (i.e.,
a dipole) in the far region. In the language of singular
perturbations, the former would be the inner region and the latter
the outer region; a complete descriptions would require the matching
of these two regions. If we disregard the outer region, we can use a
two-dimensional version of the Rayleigh-equation, which describes
the collapse of an infinite cylindrical cavity under uniform
pressure \cite{det04,ogu93,pro04}
\begin{equation}
\label{eq_ray1} \Bigg[ \frac{d(h\dot h)}{d\tau } \Bigg]\log
\frac{h}{h_\infty} + \frac{1}{2}\, {\dot h}^2 =gZ \quad .
\end{equation}
The pressure difference driving the collapse has been equated to
$\rho g Z$, where $Z$ denotes the depth below the fluid surface,
which implies that the system is composed of non-interacting
horizontal layers of fluid, with a negligible vertical velocity
component \footnote{A similar equation is used in \cite{gor05},
without the term $h_{\infty}$ and also without the hydrostatic
driving pressure $gZ$.}. Although the quantity $h_\infty$ must in
principle be determined by the matching process alluded to before,
it is expected to be of the order of a typical length scale of the
process, such as the cavity depth. Thus, strictly speaking,
$h_\infty$ is a function of time and of the Froude number. However,
near pinch-off, the time scale for the neck motion is much faster
than that for the evolution of the other parts of the cavity so that
$h_\infty$ may be considered only a function of $\mathrm{Fr}$. After
an initial expansion of the void, the collapse starts from rest at a
maximal radius $h_{max}$ (of the order of $h_{disk}$). Using this as
an initial condition, and treating $h_\infty$ as a constant, the
energy integral of Eq.~(\ref{eq_ray1}) can be readily found:
\begin{equation}
\label{eq_ray2} \left(\frac{d\tilde{h}}{d\tilde{\tau}}\right)^2 =
\frac{1}{\log (\tilde h / {\tilde{h}_\infty})}\left[1-(1/\tilde
h)^2\right] \quad ,
\end{equation}
where we have introduced the non-dimensional variables $\tilde{h}
\equiv h/h_{max}$, $\tilde{h}_\infty \equiv h_\infty/h_{max}$, and
$\tilde{\tau} \equiv \tau \sqrt{gZ/h_{max}^2}$. Close to pinch-off,
${\tilde h}^2\ll1$, such that $\tilde{h}^{-2}-1 \approx
\tilde{h}^{-2}$. With this approximation, we can integrate
Eq.~(\ref{eq_ray2}) once more to arrive at
\begin{equation}
\label{eq_ray3} 2\,\tilde \tau = \tilde{h}^2 \sqrt{\log(\tilde
h_\infty / \tilde h)} + \sqrt{\frac{\pi}{2}}\,\tilde h_\infty^2
\erfc\left(\sqrt{2\log(\tilde h_\infty / \tilde h)}\,\,\right) .
\end{equation}
For small $\tilde{h}$ the term with the complementary error function
is always small compared to the first one and their ratio vanishes
for $\tilde h \downarrow 0$. Neglecting this term we find two
asymptotic regimes
\begin{eqnarray}
\tilde{h} \left(\log (\tilde h_\infty)\right)^{1/4} &=& \sqrt{2}\,
\tilde{\tau}^{1/2}
\quad\textrm{for}\,\,  \tilde h \gg 1/{\tilde h_\infty} \,\,,\,\, \textrm{and}\label{eq_power_law1}\\
\tilde{h} \left(-\log(\tilde{h})\right)^{1/4} &=& \sqrt{2}\,
\tilde{\tau}^{1/2} \quad\textrm{for}\,\, \tilde h \ll 1/\tilde
h_\infty \,\, .\label{eq_power_law2}
\end{eqnarray}
From Eqs.~(\ref{eq_power_law1})~and~(\ref{eq_power_law2}) we
conclude that the scaling depends crucially on the value of
$h_\infty$: Initially, for the intermediate regime $h \gg
h_{max}^2/h_\infty$, we expect to find a strict power law $h \propto
\tau^{1/2}$, since $\log(\tilde{h}_\infty)$ is constant. For times
closer to the pinch-off, when $h \ll h_{max}^2/h_\infty$,
logarithmic corrections play a role, and the power law should be
modified into Eq.~(\ref{eq_power_law2}).

As $h_\infty \approx h_{max}$ in our experiments, the latter
inequality can be read as $h \ll h_{max}$, which is satisfied in
most of the region where $h(\tau)$ asymptotically behaves as a power
law (cf. Fig.~\ref{plots}a). We conclude that in our system the
logarithmic correction cannot be neglected. If we plot the quantity
$y\,=\, h(-\log{h})^{1/4}$ vs. time, we again observe a power law $y
\propto \tau^{\alpha_y}$, but with a slightly different exponent
$\alpha_y$ than the one found for $h$. In Fig.~\ref{plots}c we
compare $\alpha_h$ and $\alpha_y$ as functions of the Froude number.
As discussed before, $\alpha_h$ is substantially larger than $1/2$,
but even if the logarithmic term is included we continue to find a
slower collapse for low Froude numbers. Although the logarithmic
correction does bring the result closer to the suggested value
$1/2$, it cannot account for all of the observed deviations.

Clearly, the observed anomalous power law behavior of the neck
radius must reflect itself in the time-evolution of the free-surface
profiles of the collapsing void. If the process were self-similar,
the free-surface profiles at different times $\tau$ would superpose
when scaled by any characteristic length, e.g., the neck radius $h$.
Actually, it is found that the depth of minimum radius increases
somewhat as the collapse progresses and it is therefore necessary to
translate the profiles in the vertical direction so as to match the
position of the minimum radius point before attempting this
operation. Even if this is done, however, the results fail 
to collapse onto a single shape. The rescaled profiles become more
and more elongated as the pinch-off is approached which proves that
the collapsing void is {\it not} self-similar in a strict sense.

The free-surface shapes near the minimum point should thus not only
be characterized by $h(\tau)$, but also by a second length-scale,
the radius of curvature $R(\tau)$ in the vertical plane (see
Fig.~\ref{snapshots}e). The spatial resolution of the high-speed
camera images limits the accuracy with which $R$ can be extracted
from the experimental observations, but this quantity is easily
determined from the numerical calculations (see Fig.~\ref{plots}b).
When the radial dimensions ($H$, cf. Fig.~\ref{snapshots}e) are
scaled by $h$ and vertical ones ($Z$) by $\sqrt{hR}$, the profiles
do collapse, which may only signal that their shape is very close to
parabolic \footnote{At the minimum
$1/R(\tau)=d^2H/dZ^2|_{Z=Z_{min}}$ and the shape of the interface
can be taken to be locally parabolic, which implies $H=(\delta
Z)^2/R(\tau) + h(\tau)$ with $\delta Z=Z-Z_{min}$. The scaling of
the radial direction $H$ with $h(\tau)$ then leads to the scaling
$\sqrt{h(\tau)R(\tau)}$ for the axial direction $Z$. The aspect
ratio of the void is then given by $H/Z =
(h(\tau)/R(\tau))^{1/2}$.}. The time-evolution of this radius of
curvature is also found to follow a power law,
$R=R_0\tau^{\alpha_R}$, the exponent $\alpha_R$ of which increases
with the Froude number as can be seen in Fig.~\ref{plots}c
\footnote{The fact that both $h$ and $R$ are described by power laws
suggests that we may be dealing with self-similarity of the second
kind, in which the radial and axial coordinates are rescaled by
different power laws of time \cite{bar96}. At present there is
however insufficient experimental and theoretical ground to
substantiate such a claim.}.

The essence of the time-evolution of the void profile and the
departure from self-similarity in the strict sense is captured in
the aspect ratio $h/R$ of the collapsing void, $ h(\tau)/R(\tau) =
(h_0/R_0) \, \tau^{(\alpha_h - \alpha_R)}$, in which the prefactor
$h_0/R_0$ and the exponent $(\alpha_h - \alpha_R)$ both are found to
depend on the Froude number. It is seen in Fig.~\ref{plots}c that
$\alpha_h-\alpha_R>0$ for any finite Froude number, causing the
ratio $h(\tau)/R(\tau)$ to vanish in the limit $\tau\rightarrow0$.
This means that in this limit $R(\tau)$ becomes large with respect
to the neck radius, elongating the profiles more and more towards
the cylindrical shape close to the pinch-off, thereby justifying the
assumptions made in the derivation of Eq.~(\ref{eq_ray1}) in the
limit $\tau\rightarrow0$. A numerical fit gives
$(\alpha_h-\alpha_R)\propto\textrm{Fr}^{-0.14}$, which indicates
that $h$ and $R$ have the same time dependence as $Fr\rightarrow
\infty$ and, therefore, that self-similarity is recovered in this
limit.

A second numerical fit shows that $h_0/R_0 \propto
\mathrm{Fr}^{-0.60}$, which tends to zero as
$\mathrm{Fr}\rightarrow\infty$. This feature expresses the
experimental observation that the initial elongation of the neck is
larger for large Froude number, which effectively increases the
time-interval for which the assumption of pure radial flow is valid
[cf. Eq.~(\ref{eq_ray1})].

In conclusion, our experiments on the collapse of a giant surface
void are in excellent agreement with boundary integral calculations
without the use of any adjustable parameter. Even when we exclude
the effects of air, viscosity, and surface tension, the collapse is
found to be not self-similar in a strict sense, but governed by
power laws with non-universal, Froude-dependent exponents.
Self-similarity is recovered only in the limit of infinite Froude
number, where the influence of gravity becomes negligible and the
collapse is truly inertially driven.

\acknowledgments{This work is part of the research program of the
Stichting FOM, which is financially supported by NWO.}


\begin{thebibliography}{99}

\bibitem{ber93}
A.~L. Bertozzi, M.~P. Brenner, T.~F. Dupont, and L.~P. Kadanoff,
in {\em  Trends and Perspectives in Applied Mathematics}, edited
by L. Sirovich  (Springer, New York, 1993), Vol.~100, p.\ 155.

\bibitem{sto94}
H.~A. Stone, 
  Ann. Rev. Fluid Mech. {\bf 26},  65  (1994).

\bibitem{bre94}
M.~P. Brenner, X.~D. Shi, and S.~R. Nagel, 
Phys. Rev. Lett. {\bf 73},  3391  (1994).

\bibitem{egg97}
J. Eggers, 
Rev. Mod. Phys. {\bf 69},  865  (1997).

\bibitem{day98}
R.~F. Day, E.~J. Hinch, and J.~R. Lister, 
Phys. Rev. Lett. {\bf 80},  704  (1998).

\bibitem{zha99b}
W.~W. Zhang and J.~R. Lister, 
Phys. Rev. Lett. {\bf 83},  1151  (1999).

\bibitem{coh99}
I. Cohen, M.~P. Brenner, J. Eggers, and S.~R. Nagel, 
Phys. Rev. Lett. {\bf 83},  1147  (1999).

\bibitem{lon91} M.~S. Longuet-Higgins, B.~R. Kerman, K. Lunde,
J. Fluid Mech. {\bf 230}, 365 (1991).

\bibitem{ogu93}
H.~N. Oguz and A. Prosperetti, 
J. Fluid Mech. {\bf 257},  111  (1993).

\bibitem{dos03} P. Doshi, I. Cohen, W.~W. Zhang, M.
Siegel, P. Howell, O.~A. Basaran, S.~R. Nagel, 
Science {\bf 302}, 1185 (2003).

\bibitem{bur05} J.C. Burton, R. Waldrep, and P. Taborek,
Phys. Rev. Lett. {\bf 94}, 184502 (2005).

\bibitem{sur04} R. Suryo, P. Doshi, and O.~A. Basaran,
Phys. Fluids {\bf 16}, 4177 (2004).

\bibitem{lep03}
D. Leppinen and J.~R. Lister, 
Phys. Fluids {\bf 15},  568  (2003).

\bibitem{gor05} J.M. Gordillo, A. Sevilla, J. Rodriguez-Rodriguez,
and C. Martinez-Bazan,
Phys. Rev. Lett. {\bf 95}, 194501 (2005).

\bibitem{lat00} B.W. Zeff, B. Kleber, J. Fineberg, and D.P. Lathrop, Nature {\bf 403}, 401
(2000).

\bibitem{det04} D. Lohse, R. Bergmann, R. Mikkelsen, C. Zeilstra, D. van der Meer,
M. Versluis, K. van der Weele, M. van der Hoef, and H. Kuipers,
Phys. Rev. Lett. {\bf 93}, 198003 (2004).

\bibitem{ple77}
M.S. Plesset and A. Prosperetti, 
Annu. Rev. Fluid Mech {\bf 9}, 145 (1977)

\bibitem{hao99}
Y. Hao and A. Prosperetti, 
Phys. Fluids {\bf 11}, 1309 (1999).

\bibitem{hil96}
S. Hilgenfeldt, D. Lohse, and M.P. Brenner, 
Phys. Fluids {\bf 8}, 2808 (1996).

\bibitem{bir57} G. Birkhoff, and E. H. Zarantonello, {\em Jets, Wakes, and Cavities.}, Academic
Press, New York (1957).

\bibitem{gau98} S. Gaudet, Phys. Fluids {\bf 10}, 2489 (1998).

\bibitem{pro04}
A. Prosperetti, 
Phys. Fluids {\bf 16}, 1852 (2004).

\bibitem{bar96}
G.I. Barenblatt, {\em Scaling, self-similarity, and intermediate
asymptotics}, Cambridge University Press, Cambridge (1996).

\end{thebibliography}
\end{document}